\documentclass[conference]{IEEEtran}
\IEEEoverridecommandlockouts
\usepackage{cite}
\usepackage{amsmath,amssymb,amsfonts}
\usepackage{algorithmic}
\usepackage{graphicx}
\usepackage{textcomp}
\usepackage{xcolor}
\usepackage{subcaption}
\usepackage{booktabs}
\usepackage{xcolor} 
\usepackage{censor}
\usepackage[table]{xcolor}
\usepackage{microtype}

\PassOptionsToPackage{hyphens}{url}
\usepackage{hyperref}
\definecolor{darkblue}{rgb}{0,0,0.7}
\hypersetup{hidelinks, breaklinks, colorlinks,
  linkcolor={darkblue}, citecolor={darkblue}, urlcolor={darkblue}}

\def\BibTeX{{\rm B\kern-.05em{\sc i\kern-.025em b}\kern-.08em
    T\kern-.1667em\lower.7ex\hbox{E}\kern-.125emX}}
\begin{document}

\title{OSSMM: An Open-Source \\ Sleep Monitor and Modulator}

\author{\IEEEauthorblockN{1\textsuperscript{st} Jonny Giordano}
\IEEEauthorblockA{\textit{Hamilton Institute} \\
\textit{Maynooth University}\\
Maynooth, Ireland \\
Jonathan.Giordano.2021@mumail.ie}
\and
\IEEEauthorblockN{2\textsuperscript{nd} Fergal Stapleton}
\IEEEauthorblockA{\textit{Dept.\ of Computer Science} \\
\textit{Maynooth University}\\
Maynooth, Ireland \\
Fergal.Stapleton.2020@mumail.ie}
\and
\IEEEauthorblockN{3\textsuperscript{rd} Gabriel Palma}
\IEEEauthorblockA{\textit{School of Medicine} \\
\textit{Trinity College Dublin}\\
Dublin, Ireland \\
gpalma@tcd.ie}
\and
\IEEEauthorblockN{4\textsuperscript{th} Barak A. Pearlmutter}
\IEEEauthorblockA{\textit{Dept.\ of Computer Science} \\
\textit{Maynooth University}\\
Maynooth, Ireland \\
barak@cs.nuim.ie}
}

\maketitle

\begin{abstract}
We present the Open-Source Sleep Monitor and Modulator (OSSMM), an open-source hardware and software platform for accessible sleep research. The OSSMM comprises a small wearable headband built from 3D prints and affordable commercial-off-the-shelf (COTS) components at a material cost under €40, supported by a companion Android application. The system requires no conductive gels, disposable electrodes, or specialized equipment, and captures multiple biosignals — movement, pulse, electrooculography (EOG), and putative electroencephalography (EEG) — with wireless connectivity for data storage and potential sleep modulation capability via an onboard vibration motor. A proof-of-concept single-participant evaluation across 15 nights demonstrated that the captured biosignals support four-stage sleep classification (Wake, Light Sleep, Deep Sleep, REM) using conventional machine learning methods, with the best-performing model achieving a Macro F1-score of 0.770 and accuracy of 0.776 against a validated non-contact sleep monitor ($\kappa$=0.63 with PSG). Two technical findings are of particular note. First, inexpensive, reusable conductive thermoplastic polyurethane (CTPU) electrodes from commercial fitness chest straps captured a differential signal whose spectral properties in canonical EEG frequency bands, including signatures consistent with sleep spindles, are the principal features driving classification. Second, this signal is obtained from just two frontal electrodes without a dedicated ground reference, suggesting that practical sleep staging is achievable with simpler configurations than typically employed. All hardware designs, software, and build instructions are openly available to support replication and modification by the research community.\end{abstract}


\section{Introduction}

In 1937, Loomis, Harvey, and Hobart first revealed that sleep is not a homogeneous state, but characterized by periods of distinct neural patterns \cite{loomis1937cerebral}. This understanding was extended in 1953 with the identification of Rapid Eye Movement sleep (REM) \cite{aserinsky1953regularly}. Since then, numerous associations have been uncovered between sleep states and facets of cognition and health \cite{vogel1975rem, shepard2005history, palagini2013rem, sikka2019eeg, ketz2018closed}. The accurate identification of sleep stages is therefore essential for sleep research.

Traditionally, sleep staging has relied on manual scoring of polysomnography (PSG) by qualified experts \cite{aasm2023manual}. While considered tthe gold standard, PSG has significant trade-offs. A full PSG (Type~1) is not ecological, requiring attendance at a dedicated facility, trained staff for set-up and monitoring, while demanding the participant tolerate uncomfortable, restrictive equipment. In the United States, a single in-laboratory PSG costs on average \$1,840, while an at-home evaluation (Type~2) costs \$1,575 \cite{kim2015economic}. Many solutions have been put forth to address these drawbacks: reduced electrode placements, portable or at-home devices, and automatic sleep scoring using machine learning \cite{grozinger2002effects, miller2022validation, supratak2017deepsleepnet, toften2020validation, li2022deep}.

However, the barrier to sleep research remains high. Low-cost, accessible sleep monitoring devices perform poorly compared to PSG , while dedicated devices often cost more than a single night of PSG \cite{miller2022validation, robbins2024accuracy, schyvens2025performance}. Without significant funding, sleep researchers must choose between low-cost methods with poorer accuracies for more participants over longer periods, or high-cost methods with improved accuracies for fewer participants and shorter durations. Concerns also remain for the ease-of-use of take-home equipment, with many systems requiring prior set-up by researchers and the use of special conductive gels or electrode pads.

Active sleep monitoring, required for sleep modulation research targeting specific sleep states or neural signatures, is more demanding still. Despite promising results from sleep modulation studies, such as potential therapies for depression and cognitive investigations, progress has been limited \cite{vogel1975rem, landsness2011antidepressant, cartwright2003rem, voss2014induction}. Consider that nearly 25~years ago one sleep modulation study employed near real-time automatic sleep staging, yet more modern studies returned to manual manipulations of sleep \cite{grozinger2002effects, landsness2011antidepressant, rosales2012enhanced}.

Therefore, a solution is needed which can provide suitable sleep data collection and accurate sleep stage classification, while being affordable, transparent, and easily modifiable for varied experimental requirements. The Open-Source Sleep Monitor and Modulator (OSSMM) project responds to these needs. The platform's hardware and software are open-source, and designed for assembly with basic electronics knowledge. The system comprises a wearable electronic headband and a dedicated Android application. The headband collects movement, pulse, electrooculography (EOG), and putative electroencephalography (EEG) data and transmits data wirelessly to the connected smartphone.

In this study, we conducted a proof-of-concept technical investigation of the OSSMM platform. We hypothesized that nightly sleep data collected from the OSSMM device would support 4-stage sleep classification, namely for states "Wake," "Light Sleep," "Deep Sleep," and "REM." To test this, we employed three conventional machine learning classifiers — Support Vector Machine (SVM), Random Forest (RF), and XGBoost (XGB) — with minimal hyperparameter optimization, applied to features extracted with limited digital signal processing. Sleep stage labels were provided by a reference device, the Somnofy SM-100 by Vitalthings, a validated ($\kappa$=0.63 agreement with PSG) non-contact sleep monitor \cite{toften2020validation}. This conservative analytical approach was chosen to assess the baseline informational content of the OSSMM signals, rather than to maximize classification performance through model or feature engineering.

\section{Methods}

\subsection{Experimental Design}

A total of 17 nights of sleep data were collected using the OSSMM and Somnofy devices. The recording period was designed to yield a minimum of 14 nights of usable data, with 3 spare to accommodate unforeseen disruptions (e.g. illness, equipment faults). This technical evaluation employed a single-participant design (N=1) with one researcher serving as the sole participant, yielding 15,285 30-second epochs (i.e. observations) across 15 qualifying nights. The single-participant design was chosen as this study constitutes a technical validation of the platform, assessing system functionality and hardware reliability prior to multi-participant validation. This experimental design choice aligns with the IEEE Code of Ethics, avoiding unnecessary risk to participants ahead of full system validation \cite{ieee_code_of_ethics_2020}. Generalizability to a broader population is addressed as a limitation in Section~IV.

The nightly protocol was as follows: The CTPU (conductive Thermoplastic Polyurethane) electrodes were wiped clean with a damp cloth and dried to remove any oils that may have accumulated from the previous night. The device was donned, powered on, and connected to the companion application. A Google Pixel~9 smartphone running Android~16 ran V1.0.4 of the OSSMM application. The application's real-time plots were used to verify that movement, pulse, and EOG data were being detected and recorded. A typical bedtime activity (e.g., reading) was then carried out for at least 15~minutes to augment the number of Wake observations. Immediately before attempting sleep, a modified consensus sleep diary was partially completed before sleep \cite{carney2012consensus}.

Upon waking the following morning, data recording was halted, and the headband powered off and removed. The headband was connected to a USB-C charger and the remaining sections of the consensus sleep diary were completed covering nocturnal awakenings, perceived sleep quality, and the headband displacement questions described below.

The sleep diary was augmented with two self-report questions: ``Did you notice the headband move during your sleep?'' and ``Did the headband come off during sleep?'' These questions were included to inform iterative improvements to the headband design and to monitor data quality impacts from potential device displacement during recording. If the device shifted position or came off completely during the night, its position was to be restored without verification of proper signal capture. 

The SM-100 reference device was placed on a bedside table with its distance setting configured appropriately through the Somnofy companion application. Once powered on and configured, the SM-100 continuously collected data without requiring daily initiation for recording. Data collected by the SM-100 were retrieved from the Somnofy server via an Application Programming Interface (API) provided by VitalThings. Each classified SM-100 observation captured approximately 30~seconds of data, with epochs labeled as ``Wake,'' ``Light Sleep,'' ``Deep Sleep,'' ``REM,'' or ``Not Detected.''

\subsection{Data Methods}
For analysis, raw data from the OSSMM device was time locked to the 30~second bins of SM-100 data which provided classification labels. Features were derived from the OSSMM data for these 30~second epochs for machine learning classification.

Of the 17~nights, 2 were discarded due to mains power interruptions (Nights~2 and~10), which prevented the SM-100 device from achieving full nightly recording. Of the remaining 15~nights, three (Nights~3, 13, and~17) were randomly withheld for final testing using a random number generator. Test nights were not allowed to be consecutive in order to promote longitudinal representation of the data.

\subsection{OSSMM System Architecture}

\begin{table*}[!t]
\caption{Hardware Component Specification and Cost Analysis}
\label{tab:components}
\renewcommand{\arraystretch}{1.2}
\begin{center}
\begin{tabular}{p{3.5cm}p{6.5cm}p{1cm}p{1.5cm}p{1.5cm}}
\toprule
\textbf{Component} & \textbf{Product Specification} & \textbf{Qty} & \textbf{Unit Cost} & \textbf{Total Cost} \\
\midrule
\multicolumn{5}{l}{\textit{Electronics}} \\
\rowcolor{gray!10} Microcontroller & Seeed Xiao nRF52840 Sense & 1 & €{18.20} & €{18.20} \\
 EOG Amplifier & AD8232 PCB & 1 & ~€{2.20} & ~€{2.20} \\
\rowcolor{gray!10} PPG Sensor & PulseSensor Clone & 1 & ~€{2.20} & ~€{2.20} \\
 Vibration Motor & Vibration Motor Board & 1 & ~€{3.00} & ~€{3.00} \\
\rowcolor{gray!10} Battery & 3.7V LiPo 200 mAh & 1 & ~€{7.80} & ~€{7.80} \\
\midrule
\multicolumn{5}{l}{\textit{Materials}} \\
\rowcolor{gray!10} CTPU Electrodes & Heart Rate Monitor Chest Strap & 1 & ~€{2.70} & ~€{2.70} \\
 Electronic Casing & Prusament PLA & 14.51\,g & €{0.03745}/g & ~€{0.60} \\
\rowcolor{gray!10} Housing Attachment & Siraya Tech TPU-85A & 29.01\,g & €{0.03755}/g & ~€{1.10} \\
\midrule
\multicolumn{4}{r}{\textbf{Total Cost}} & \textbf{€{37.80}} \\
\bottomrule
\multicolumn{5}{p{15cm}}{\textit{Note: Solder, wiring, and snap fasteners are excluded from this cost analysis. Prices have been rounded up to the nearest €0.10 and include shipping.}}
\end{tabular}
\end{center}
\end{table*}

A major motivation behind the creation of the OSSMM system was to establish a platform which researchers themselves could assemble in bulk, modify as desired, dispense to numerous participants for use over prolonged periods, and later reuse. Therefore the design project prioritized, second only to safety, ease of construction, accessibility of parts, cost, and ergonomics.

\begin{figure}[htbp]
\centerline{\includegraphics[width=0.45\textwidth]{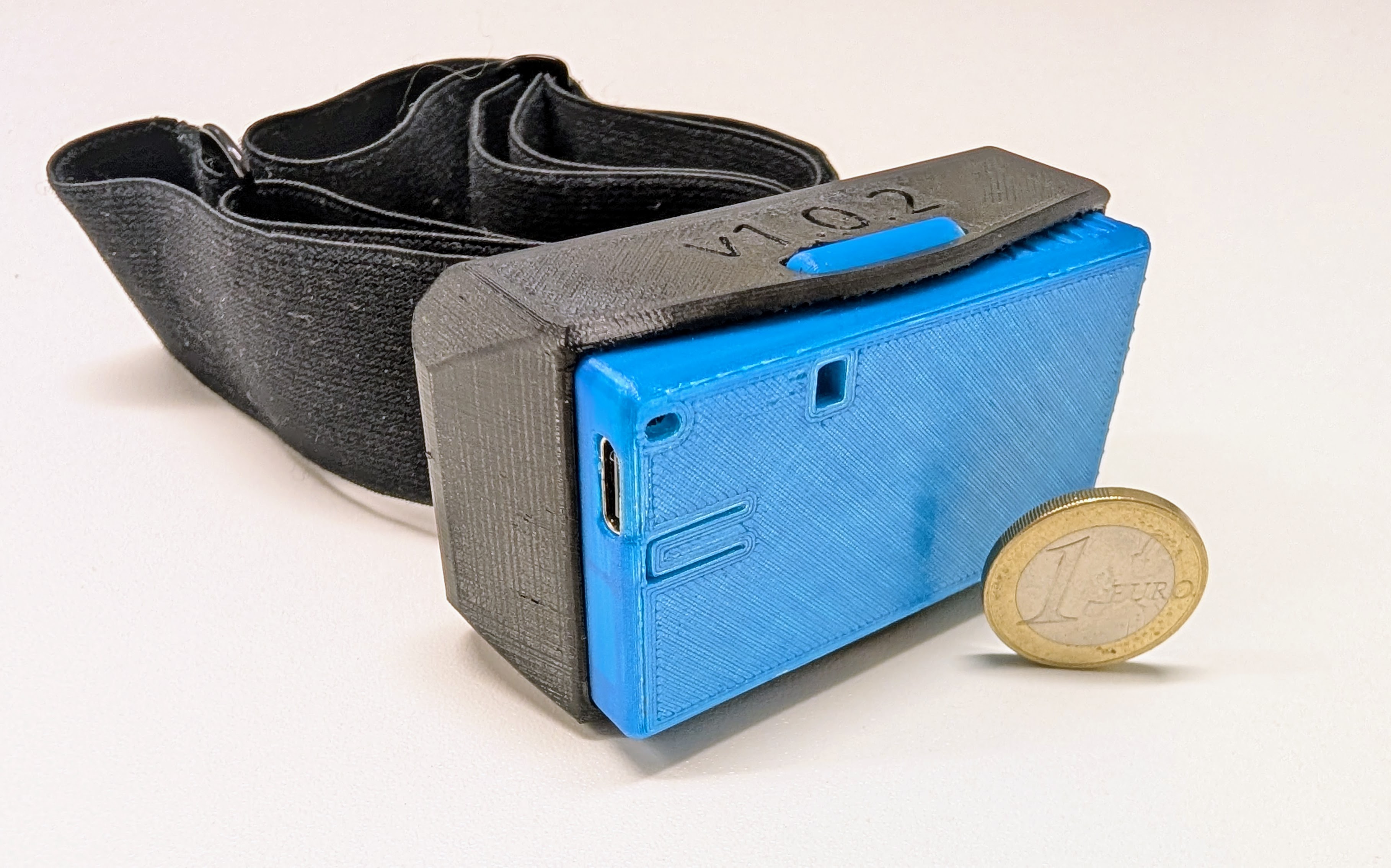}}
\caption{OSSMM headband compared with €1 coin.}
\label{ossmm_headband}
\end{figure}

A major challenge in sleep research is collecting data in an ecological manner, minimizing disturbance during measurement. Size and felt-weight were therefore minimized as much as possible. Figure~\ref{ossmm_headband} demonstrates the headband in comparison with the size of a €1 coin. The bounding dimensions of the headband, excluding the strap, are 79.1 × 47.2 × 31.0~mm, and a unit weighs 76.6~g with a 200~mAh battery.

To address cost and accessibility, the headband employs budget commercial-off-the-shelf (COTS) components available at hobbyist electronic suppliers or general marketplaces. Table~\ref{tab:components} shows a full cost breakdown. By offloading storage, chronometry, and user interface demands to an Android device through the companion application, the hardware complexity and cost were significantly reduced, keeping the total parts cost under €40 at 2024 pricing. Because of the modularity of the platform, components can be excluded if they do not pertain to a researcher's particular interest, thereby lowering the cost further.

The platform was designed for simple fabrication and assembly, using the fewest components necessary with minimal modification. Fabrication requires a 3D printer, soldering kit, and a computer. All components can be purchased from major retailers; 3D prints can also be ordered from vendors using the available OSSMM print files.

\begin{figure}[htbp]
\centerline{\includegraphics[width=0.45\textwidth]{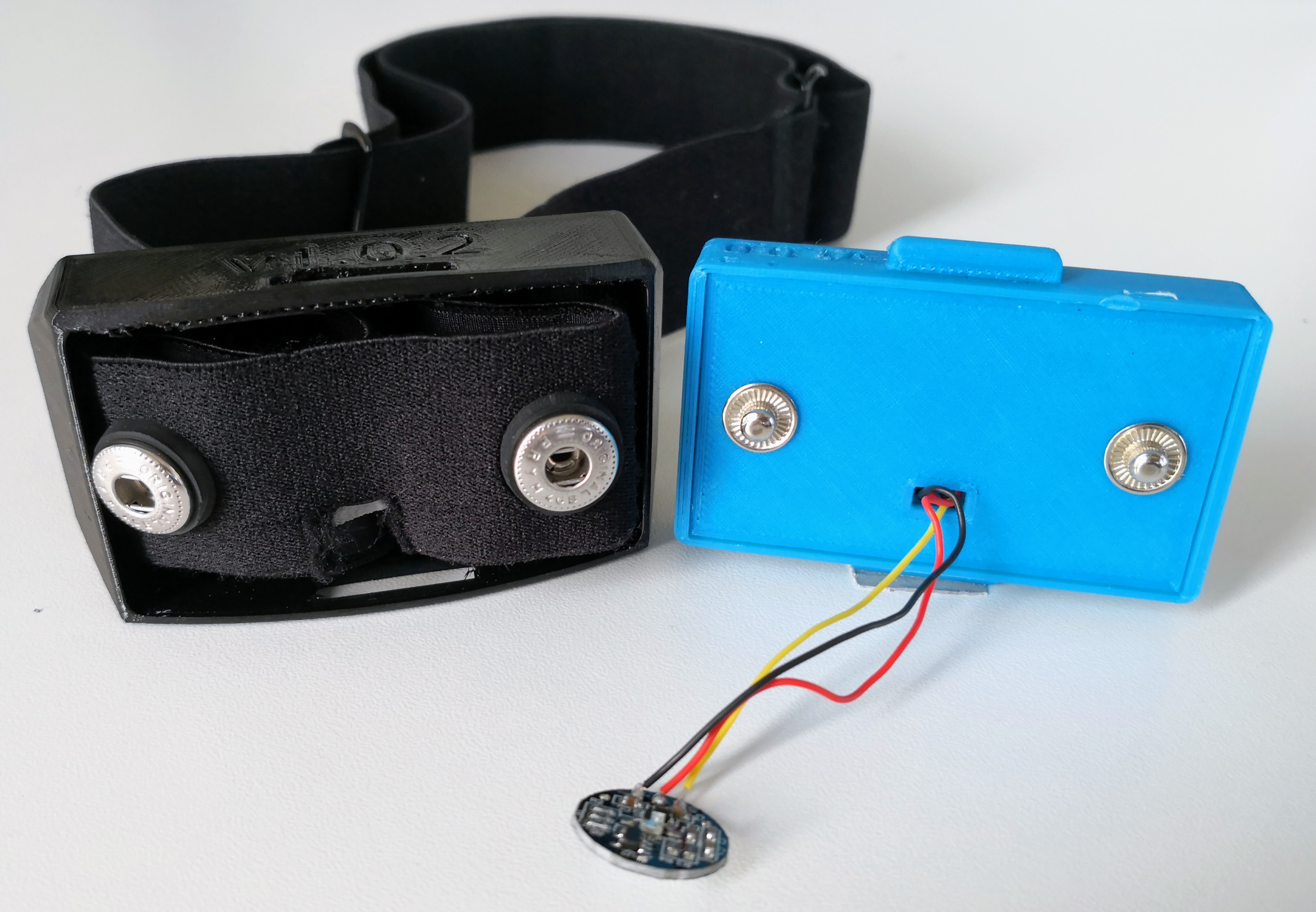}}
\caption{The OSSMM Headband with interchangeable parts disconnected for viewing.}
\label{ossmm_headband_open}
\end{figure}

As seen in Figure~\ref{ossmm_headband_open}, the headband consists of three principal components: the strap, the receiver (here printed in black filament), and the electronic case (here printed in blue filament). These major components allow for quick interchange, permitting rapid fixes or cleaning, such as when transferring a device between participants. The components connect via locking tabs and slots or metal snap-fasteners, the latter also serving as connection points to transfer EOG/EEG signals to the AD8232 inputs.

The OSSMM headband comprises just 11 components excluding wiring (Table~\ref{tab:components}). A Seeed Xiao nRF52840 Sense MCU coordinates data acquisition from three dedicated sensor boards: an IMU collecting 6~DoF movement data, a PulseSensor clone for PPG, and an AD8232 EKG board in ``Cardiac Monitor'' configuration for EOG and EEG acquisition, using a differential amplifier with 0.5~Hz high-pass and 40~Hz low-pass filters \cite{nRF52840_product_spec, PulseSensor_datasheet, AD8232_datasheet, Sparkfun_AD8232_schematic}. Notably, the right leg drive amplifier is disconnected, and no dedicated ground reference electrode is employed. EEG/EOG signals are captured via two CTPU electrodes embedded in a heart-rate monitor chest strap, an inexpensive and commonly available consumer fitness band. Electrode placement acquires potential differences approximately equivalent to $\text{Avg(F7+Fp1)} - \text{Avg(Fp2+F8)}$ in the 10-20 system \cite{jasper1958ten}. A vibration motor board is also included for potential sleep modulation.

The MCU is programmed using Arduino with the Seeed nRF52 V1.1.8 board manager. The OSSMM V1.0.4 Android application was written using Dart and the Flutter framework. The MCU employs Bluetooth Low Energy (BLE) to wirelessly transmit collected data to the companion application through an AES-128 encrypted channel. In a strictly sleep monitoring role, the 200~mAh battery lasts over 23 hours of continuous use. An MCU-mounted PDM MEMS microphone exists in the hardware but is not yet implemented in software.

The greatest priority in OSSMM design and construction was safety. Filaments which make skin contact met dermatological safety standards \cite{Kerbio2024_Irritation, Kerbio2024_Cytotoxicity, Kerbio2024_Sensitization}. Importantly, the OSSMM headband has no exposed wiring and delivers no current to the user.

All code, hardware designs, and an instructional build guide have been available in a dedicated repository since May 2025 at: \url{https://github.com/jvgiordano/OSSMM-Version-1}.


\subsection{Signal Processing and Feature Selection}
All sensor data were digitized at 250~Hz with 10-bit ADC resolution. IMU data underwent baseline shifting to ensure positive values for optimized Bluetooth Low Energy (BLE) transmission. No further signal processing occurred on the MCU. The classification models utilized 42 features.

\subsubsection{Pulse Features}
Pulse data contributed 8 features, principally capturing heart rate and heart rate variability through frequency domain and harmonic-based detection approaches. Signal preprocessing consisted of a 4th-order Butterworth low-pass filter with a 10~Hz cutoff frequency. For frequency-domain analysis, pulse data were additionally preprocessed by mean-subtraction detrending followed by application of a Hamming window to address spectral leakage.

\subsubsection{IMU Features}
IMU data provided 12 features exclusively characterizing movement rather than positional characteristics. Positional characteristics were avoided to improve generalizability as positional correlations with sleep states might be specific to the limited participant sample. Movement quantification consisted of three-dimensional magnitude calculations, basic statistical parameters, and adaptive statistical thresholding to identify sudden movement changes.

\subsubsection{EOG and EEG Features}
Finally, EOG/EEG data contributed 22 features. Fluctuations in impedance due to compression or movement of electrodes, and therefore unwanted signal changes, are common in wearable bio-signal recording, particularly EEG. To differentiate ocular movement from such motion artifacts, several adaptive thresholding features were employed. Basic statistical parameters such as standard deviation (STD) and Interquartile Range (IQR) were also calculated.

Although the headband was not originally designed for EEG capture, five power features in neurophysiologically relevant ranges were calculated: delta (\textless4~Hz), theta (4--8~Hz), alpha (8--13~Hz), beta (13--30~Hz), and gamma (30--100~Hz). Spectral density estimation utilized periodograms with a Tukey window (10\% cosine taper).

\subsection{Classifiers}
Three classic machine learning classifiers were examined in a comparative analysis: SVM, RF, and XGB \cite{pedregosa2011scikit, chen2016xgboost}. Hyperparameter selection was conducted by manual search, in which configurations were purposefully designed to explore the breadth of each model's hyperparameter space. Configurations were selected to reflect meaningful trade-offs; for example, tree-based ensemble methods varied tree depth and number of estimators jointly to assess the effect of model complexity on classification performance.

Hyperparameter evaluation was carried out on the 12 training nights using Group K-Fold cross-validation with nights as the grouping unit (6 folds, 10 training nights and 2 validation nights per fold). Night-level grouping was employed to prevent within-night data leakage: consecutive epochs within the same night share highly correlated signal characteristics due to the gradual nature of sleep state transitions, and splitting these across training and validation folds would artificially inflate performance estimates. While between-night dependencies in sleep architecture may exist (e.g., compensatory sleep following a poor night's sleep), the classifiers evaluate each 30-second epoch based solely on its summary features, without access to information from other epochs or nights; temporal ordering between nights was therefore not applicable. Five configurations per classifier were tested with the goal of validating signal quality and system design rather than maximizing classifier performance. 

One configuration for each classifier (i.e. SVM, RF, XGB) was selected for testing on the withheld data. The selection criterion was the highest Macro F1-score, with accuracy used as a tiebreaker for configurations achieving the same F1-score. For SVM, regularization parameter C was varied across three orders of magnitude (0.01--10.0) and gamma across both scale-based and fixed values. For Random Forest, the number of estimators ranged from 50 to 500 and maximum depth from 6 to unconstrained. For XGBoost, learning rates ranged from 0.05 to 0.1, estimators from 50 to 500, and maximum depth from 4 to~10.

The best-performing configuration from each classifier was then trained on all 12 training nights and tested on the three withheld nights. Class balancing was handled using Synthetic Minority Over-Sampling Technique (SMOTE) during training \cite{chawla2002smote}.

\section{Results}

\subsection{Data}

From 15 nights of sleep data, 15,285 30-second epochs qualified for analysis. 40 epochs were excluded due to insufficient data samples (\textless95\% of expected samples), all occurring during BLE connection initialization or shutdown. These boundary epochs reflect the stabilization period of the wireless connection or disconnection. Additionally, 10 epochs were removed due to a Somnofy device classification of `Not Detected'.

\begin{table}[h]
\caption{Sleep stage distribution in training and test sets}
\label{tab:dataset_distribution}
\renewcommand{\arraystretch}{1.3}
\begin{center}
\begin{tabular}{p{2.4cm}cc}
\toprule
\textbf{Sleep Stage} & \textbf{Training Set (\%)} & \textbf{Test Set (\%)} \\
\midrule
Deep Sleep & 14.93 & 16.64 \\
Light Sleep & 48.28 & 45.91 \\
REM & 23.58 & 21.65 \\
Wake & 13.21 & 15.80 \\
\midrule
\textbf{Total Epochs} & \textbf{12,449} & \textbf{2,836} \\
\bottomrule
\end{tabular}
\end{center}
\end{table}

The dataset was split into 12~nights for training (12,449~epochs, 81.45\% of total) and 3~nights for testing (2,836~epochs, 18.55\% of total). Table~\ref{tab:dataset_distribution} shows the class distribution.

\begin{figure}[h]
\centerline{\includegraphics[width=0.5\textwidth]{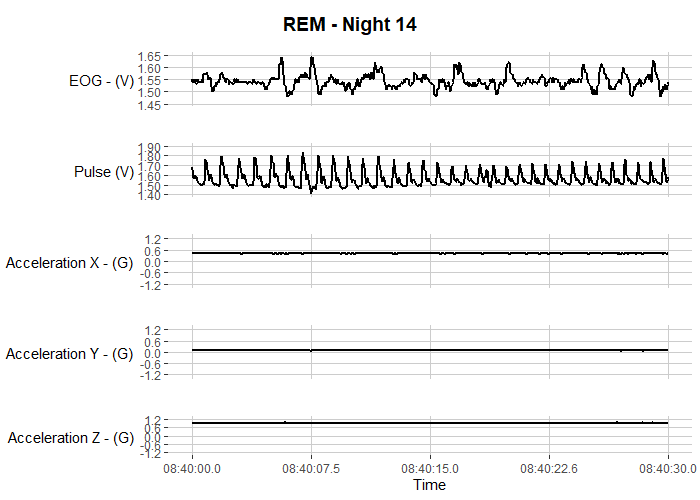}}
\caption{REM observation in Night 14 capturing portions of Epoch 1002 and 1003. Saccades are observed in the EOG plot, with no head motion observed by the accelerometer.}
\label{sample_observation}
\vspace{-8pt}
\end{figure}

Figure~\ref{sample_observation} shows a 30-second epoch of raw data collected from the sensors EOG, Pulse, and Accelerometer Sensors. Indeed, significant eye movement is observed within the EOG channel. Heart rate during the first half of the observation is approximately 64 BPM, increasing to 72 BPM in the later half. This increased heart rate variability is indicative of REM sleep. No head movement is indicated by the accelerometer axes. The sleep diary notes
waking from a dream, with wake onset occurring approximately 10 minutes after this epoch.

\begin{figure}[htbp]
\centerline{\includegraphics[width=0.5\textwidth]{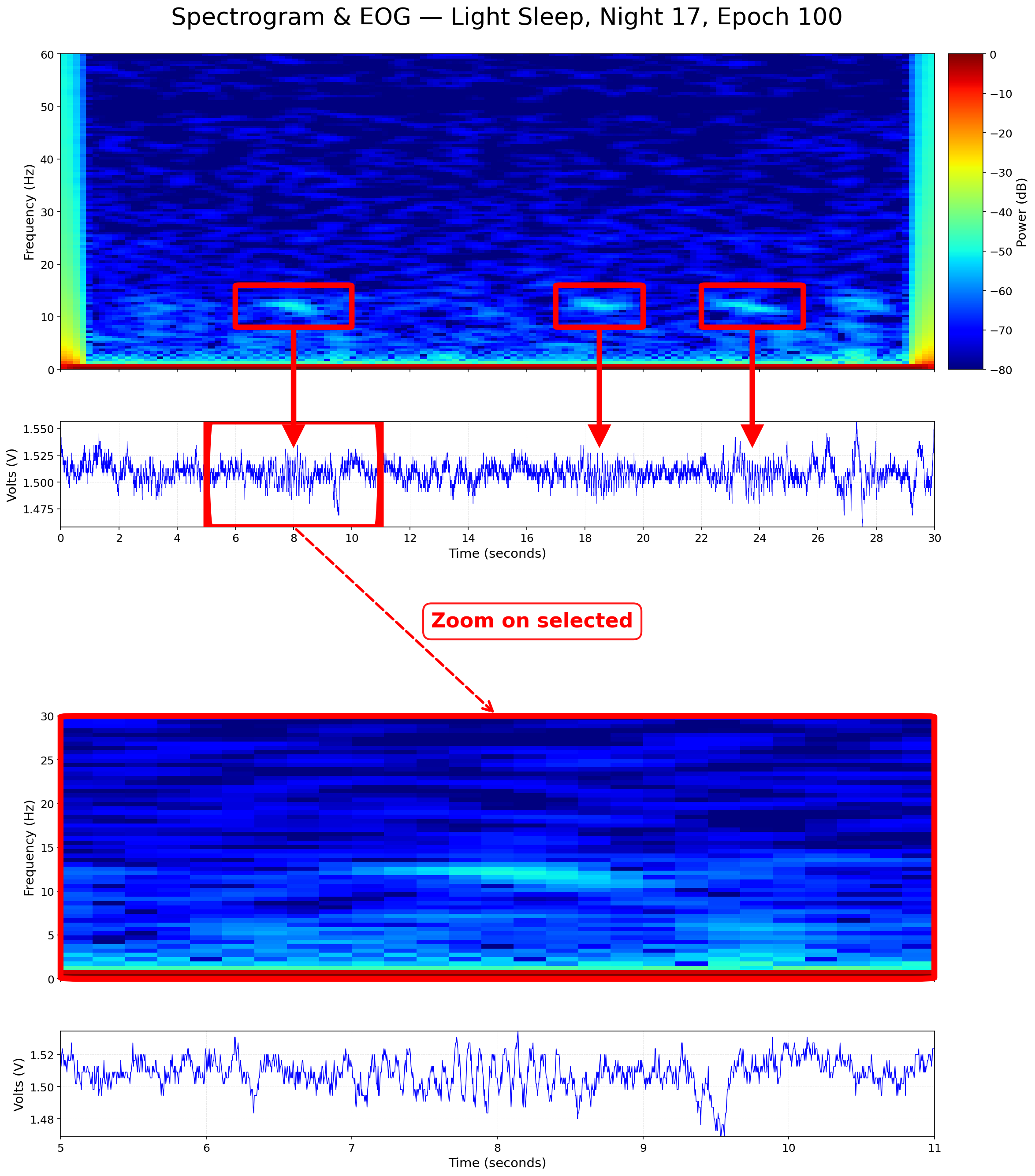}}
\caption{Spectrogram and time-series plot of the EOG channel on Night 17, Epoch 100. The upper spectrogram (0–60~Hz) spans the full 30-second epoch, with three sleep spindle–like signatures highlighted (8–16~Hz). The lower panels show a zoomed view (0–30~Hz, 5–11~s) of the first signature.}
\label{spectrogram}
\end{figure}

Figure~\ref{spectrogram} shows the spectrogram and associated time-series plot of the EOG/EEG channel for a Light Sleep observation as denoted by the SM-100. Three signatures matching canonical sleep spindles are highlighted in the spectrogram, with corresponding arrows to the location in the time-series plot below.

\subsection{Classifier Configuration Selection}

Five configurations per classifier were evaluated using 6-fold Group K-Fold cross-validation on the training data, with configurations designed to explore meaningful trade-offs across each model's hyperparameter space (e.g., jointly varying tree depth and number of estimators for tree-based ensemble methods). Table~\ref{tab:selected_configs} shows the best-performing configuration for each classifier, selected by highest Macro F1-score with accuracy as a tiebreaker.

\begin{table}[h]
\caption{Selected Hyperparameter Configurations with Cross-Validation Performance}
\label{tab:selected_configs}
\renewcommand{\arraystretch}{1.3}
\begin{center}
\begin{tabular}{p{1.1cm}p{4.6cm}p{0.6cm}p{0.6cm}}
\toprule
\textbf{Model} & \textbf{Hyperparameters} & \textbf{Macro-F1} & \textbf{Acc.} \\
\midrule
SVM & C=1.0, gamma=scale, kernel=rbf & 0.551 & 0.548 \\
RF & \raggedright n\_est.=150, max\_d.=15, min\_split=3, min\_leaf=2 & 0.766 & 0.779 \\
XGB & \raggedright max\_d.=6, lr=0.1, n\_est.=100, subsamp.=0.8, colsamp.=0.8 & 0.762 & 0.776 \\
\bottomrule
\end{tabular}
\end{center}
\vspace{-8pt}
\end{table}

For SVM, the default baseline configuration performed best, though all five configurations spanned an F1 range of 0.072 (0.479--0.551). Random Forest and XGBoost were notably insensitive to hyperparameter choices, with F1 ranges of just 0.028 (0.738--0.766) and 0.014 (0.748--0.762) respectively. This stability suggests that the classification performance is driven primarily by informational content of the captured signals rather than model tuning.

\subsection{Classifier Performance on Test Data}

\begin{figure}[htbp]
\centerline{\includegraphics[width=0.5\textwidth]{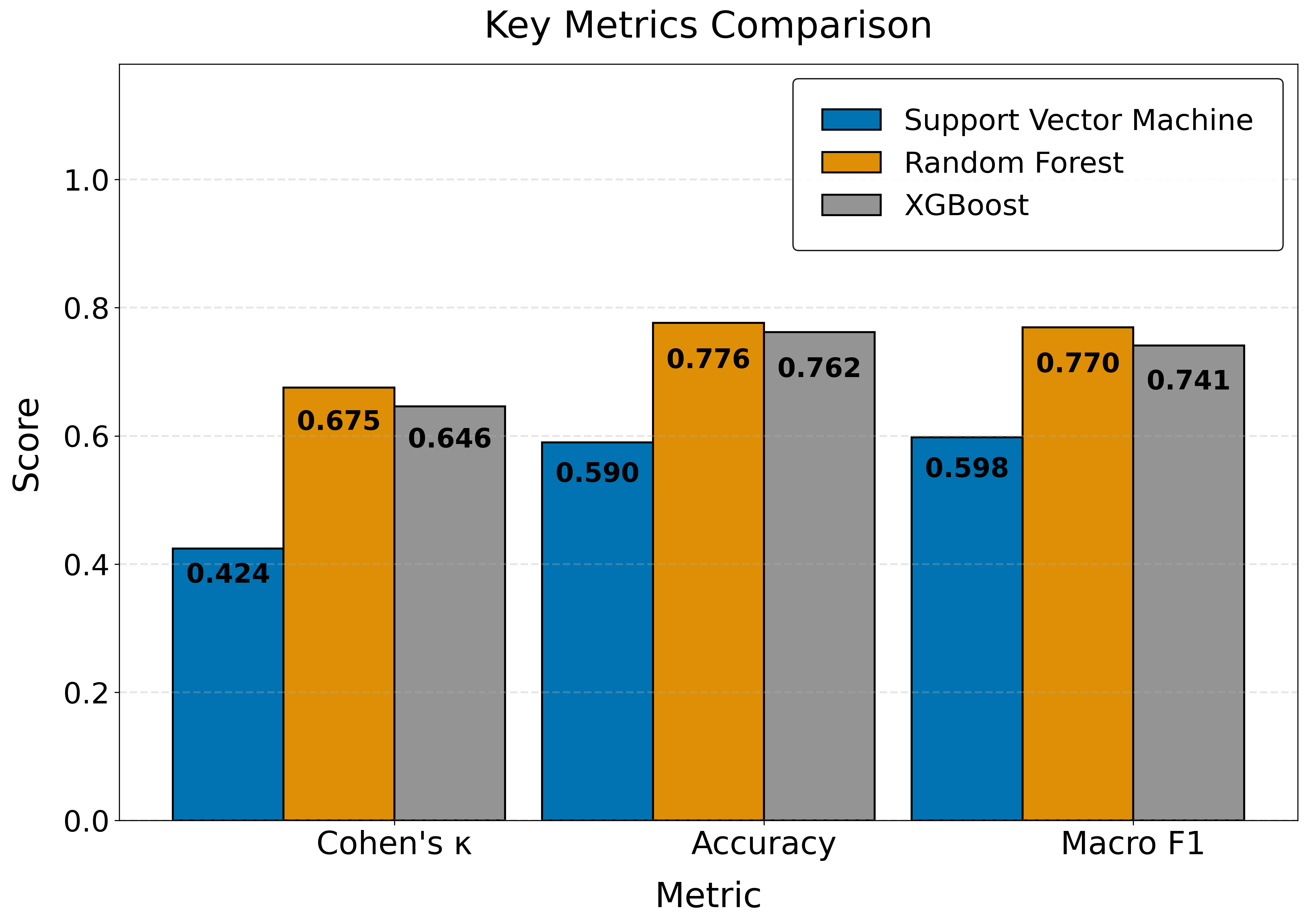}}
\caption{Classification performance between SVM, RF, and XGB on withheld data.}
\vspace{-8pt}
\label{model_comparison}
\end{figure}

The final evaluation on the withheld test data, shown in Figure~\ref{model_comparison}, revealed a clear performance hierarchy. The SVM model was the least effective, yielding an F1-score of 0.598 and accuracy of 0.590. In contrast, both Random Forest and XGBoost demonstrated strong and marginally different performances. The XGBoost model achieved an F1-Score of 0.741 and accuracy of 0.762. Random Forest obtained the highest performance on both metrics, with an F1-Score of 0.770 and an accuracy of 0.776.

\begin{figure*}[htbp]
    \centering
    \begin{subfigure}{0.46\textwidth}
        \centering
        \includegraphics[width=\textwidth]{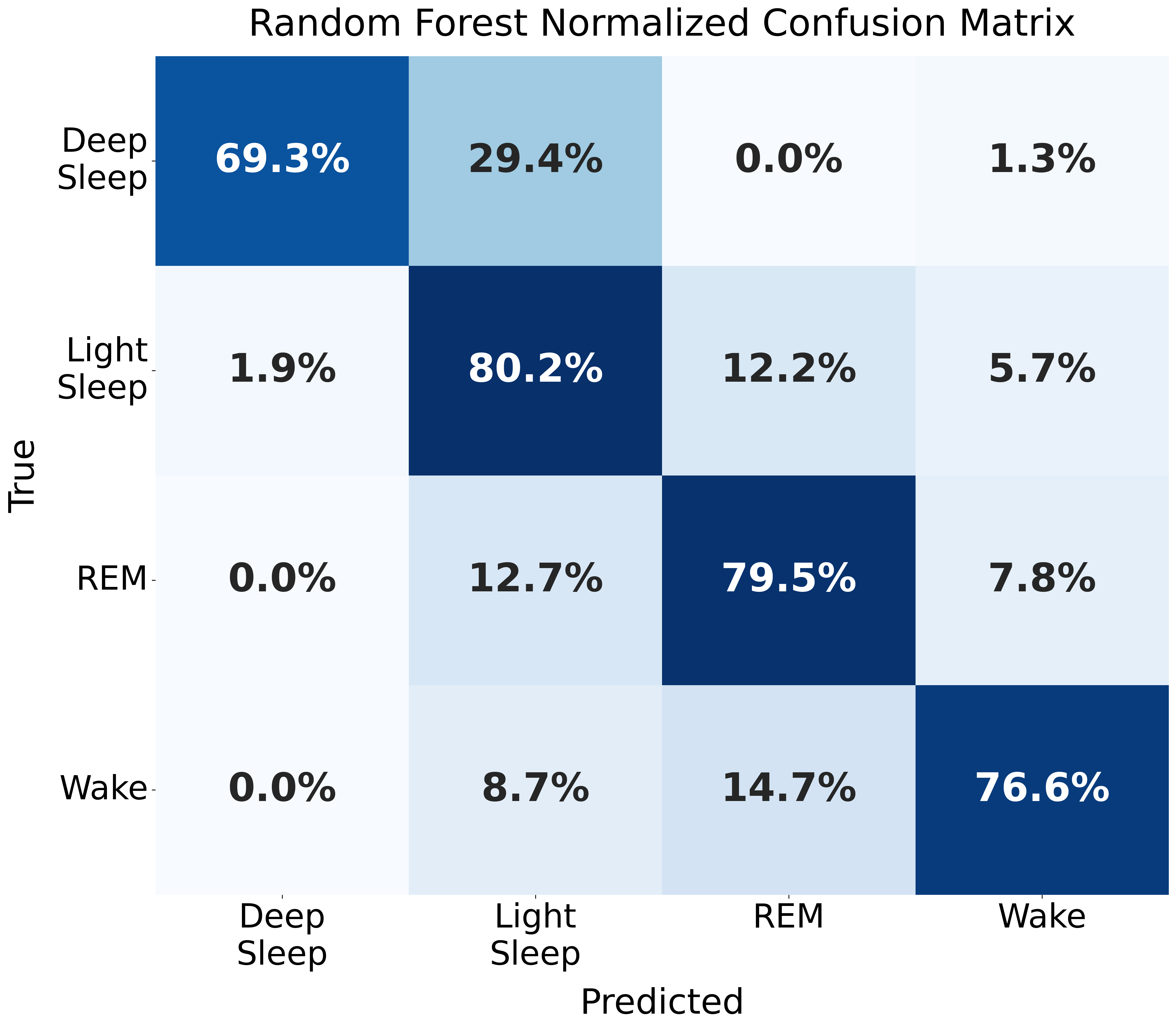}
        \caption{Normalized Confusion Matrix}
        \label{fig:rf_confusion_matrix}
    \end{subfigure}
    \hfill
    \begin{subfigure}{0.50\textwidth}
        \centering
        \includegraphics[width=\textwidth]{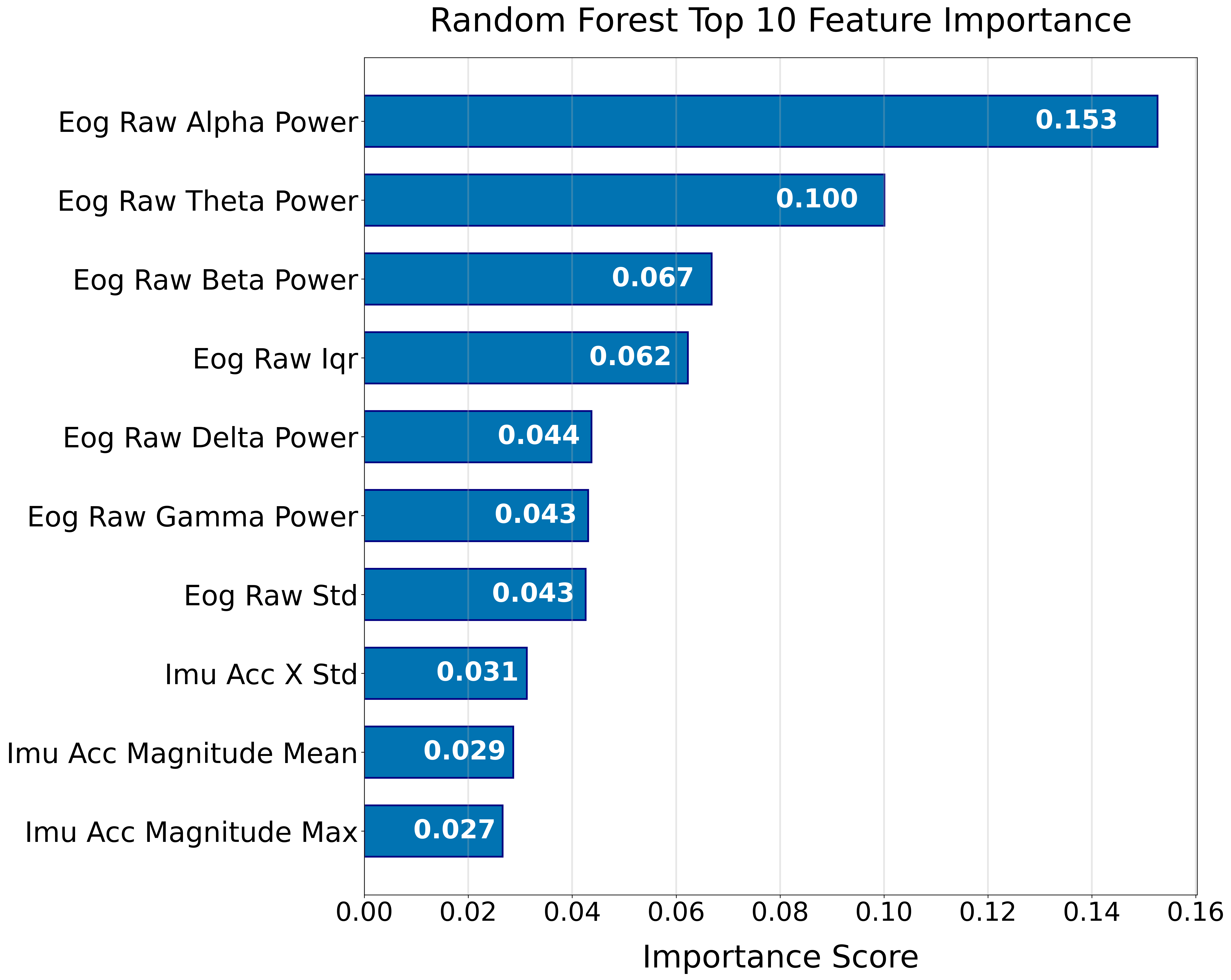}
        \caption{Top 10 Feature Importance (Mean Decrease in Impurity)}
        \label{fig:rf_feature_importance}
    \end{subfigure}
    \caption{Random Forest Classifier: (a) Normalized confusion matrix on withheld test data, and (b) top 10 feature importance rankings. EOG/EEG-derived features, particularly alpha power, dominate the importance rankings.}
    \label{fig:rf_combined}
\end{figure*}

Figure~\ref{fig:rf_confusion_matrix} shows the normalized confusion matrix for the best-performing Random Forest classifier. The model achieved strong performance across all sleep-wake states, with a clear distance gradient in misclassifications whereby adjacent sleep stages were more frequently confused than distant stages. For example, the model never misclassified REM or Wake as Deep Sleep, and similarly never misclassified Deep Sleep as REM. Deep Sleep was misclassified as Wake just 1.3\% of the time. XGBoost exhibited a similar pattern, but with notably lower Deep Sleep recall, misclassifying 47.0\% of Deep Sleep epochs as Light Sleep compared to 29.4\% for Random Forest.

All classifiers demonstrated difficulty with Deep Sleep. Specificity for Deep Sleep was high, but recall was low, with Deep Sleep often misclassified as Light Sleep. The SVM classifier struggled particularly with REM versus Light Sleep differentiation, and over-predicted Wake, misclassifying other sleep stages as Wake more than 11\% of the time.

Both Random Forest and XGBoost provide feature importance scores, and these were calculated with default settings. The top 10 features for Random Forest are shown in Figure~\ref{fig:rf_feature_importance}. The two models demonstrated strong agreement in feature importance rankings. EOG/EEG-derived features dominated both models' top 10 lists, with all five EEG frequency band power features (alpha, beta, theta, gamma, and delta) appearing in both rankings. Both models identified alpha power as the most important feature, with Random Forest assigning it an importance of 0.153 and XGBoost 0.148. EOG IQR (the interquartile range of EOG values), also ranked highly in both models, at fourth in Random Forest and second in XGBoost.

IMU-derived features appeared in both top 10 lists, though with lower importance scores, while pulse-derived features were absent from the top rankings in both models.

\section{Discussion}

\subsection{Hypothesis and Results}
We hypothesized that data collected by the OSSMM platform could be used to classify the sleep stage labels produced by a Somnofy SM-100 device for the same nightly observations. The confusion matrix patterns for Random Forest and XGBoost support this. The classifiers more commonly mistook sleep-wake states which are more similar by definition, while rarely mismatching those most dissimilar. Both models showed some Wake-to-Light Sleep misclassification; however, this distinction is considered one of the most difficult in sleep staging.

Given the imbalanced test set, a classifier performing at stratified chance would achieve an accuracy of 31.0\% and one always predicting the most numerous class, Light Sleep, would achieve 45.9\%. That the Random Forest classifier achieved an accuracy of 77.6\% with a Macro F1-score of 77.0\% can be considered practically higher than these baselines. Therefore, we consider the hypothesis that data collected by OSSMM could be used to classify sleep states with Somnofy labels supported. Importantly, this validation assesses whether the OSSMM signals contain sufficient information to characterize the current sleep state — the classifiers evaluate each epoch independently, without access to sequential or temporal context.

\subsection{Notable Findings}

Concerning affordability and accessibility, the results support the feasibility of fabricating a low-cost sleep monitor (\textless€40) which can successfully classify sleep stages as determined by a validated device using easily available COTS components and a typical 3D printer. That this was accomplished without significant signal processing, feature extraction or selection, or extensive hyperparameter refinements attests to the informational content of the captured signals.

In consideration for usage by researchers and participants, the design of the OSSMM platform shows that special conductive gels, disposable electrodes, or complex user instructions are not required. The OSSMM headband permits repeated use, in this case proven over the course of more than 2 weeks, without any special maintenance aside from a simple wipe down cleaning and daily charging via USB-C charger.

Concerning technical advances, there are two of principal interest. First, our findings indicated that the most important features for sleep state determination were spectral power features in canonical EEG frequency bands. While this may have been expected given that EEG signatures are the predominant defining features of sleep stages, it was unexpected that a signal with such neurophysiologically consistent properties would be captured through the CTPU electrodes of a heart-rate monitor chest strap and the selected filtering and amplifier circuit. A visual investigation with spectrograms, as shown in Figure~\ref{spectrogram}, revealed signatures ($\approx$12~Hz, $\approx$0.5--2~seconds) matching canonical sleep spindles (11--16~Hz, 0.5--2~seconds), occurring predominantly in Light Sleep observations \cite{li2022deep, NayakAnilkumar_EEGNormalSleep_2023}. While formal validation against gold-standard PSG is required to confirm the origin and quality of these signals, the spectral and temporal characteristics are strongly consistent with known EEG phenomena. This is of particular note because it suggests that these straps and respective electrodes may be suitable for EEG-related research, having the benefits of being commonly available, affordable, reusable, and washable.

Second, and most importantly, our results demonstrate that the differential signal captured from just two frontal electrodes, without a reference ground electrode as is standard for EEG data capture, contains information sufficient for practical sleep stage classification. If, as the spectral and temporal evidence in the preceding paragraph suggests, this signal is substantially cortical in origin (i.e. EEG), this finding aligns with that of Lucey et al.\, who demonstrated that a bipolar frontal inter-hemispheric EEG derivation achieved substantial agreement with polysomnography ( $\kappa$= 0.67) for sleep stage classification \cite{lucey2016comparison}. Tashakori et al.\ reached an apparently contradictory conclusion, finding that hemispheric differences were not clinically meaningful \cite{tashakori2024interhemispheric}. However, this may reflect a methodological distinction: Tashakori et al.\ evaluated hemispheric differences through statistical comparison of signal features, rather than directly testing whether the difference signal itself could serve as an input for classification. Our findings support the feasibility of simpler sleep monitoring configurations using fewer electrodes and no dedicated ground reference.

\subsection{Limitations and Future Work}

This experiment had two principal limitations. First, the single-participant design limits generalizability, as the classifiers may have targeted idiosyncratic characteristics of the participant. Second, while validated, the Somnofy SM-100 is not the gold standard for sleep measurement and uses biological correlates (e.g., respiration rate) for sleep staging. This study is therefore a technical validation against Somnofy reference labels, rather than true sleep states as determined by PSG.

One minor limitation is that while the sleep diary captured the frequency and duration of nocturnal awakenings, it did not record the nature of these awakenings (e.g., getting water), which may have implications for signal quality assessment.

Future investigations with a suitable and large participant pool and PSGs serving as reference measures will allow for performance assessment against true sleep states. Formal assessment of the differential frontal signal against gold-standard PSG is needed to characterize signal quality and confirm the cortical origin of the spectral features identified in this study. While the classifiers here evaluated each epoch independently based on summary features, future work could employ time-series cross-validation to explore if temporal ordering of recording sessions influences performance estimates. Furthermore, investigations into advanced digital signal processing, better feature selection, and refined machine learning classifiers are likely to result in improved performance.

\section{Conclusion}

This work presented the Open-Source Sleep Monitor and Modulator (OSSMM), a fully transparent, low-cost sleep monitoring platform built from commonly available components for under €40. Using conventional machine learning classifiers with minimal optimization, the OSSMM achieved 4-stage sleep classification with a Macro F1-score of 0.770 and accuracy of 0.776 against a validated reference device. 

Two technical findings are of particular note: first, that inexpensive, reusable CTPU electrodes from commercial fitness chest straps capture a differential signal with spectral properties in canonical EEG frequency bands that are strongly consistent with known EEG phenomena and contribute substantially to sleep stage classification; and second, that this differential signal, captured from just two frontal electrodes without a reference ground, contains information sufficient for practical sleep staging. Together, these findings suggest that the barrier to entry for sleep research can be substantially lowered without a significant sacrifice to classification performance. 

All hardware designs, software, and build instructions are openly available to support replication, modification, and extension by the research community.

\textbf{Ethics} This study received Tier 3 ethics approval from the Biomedical \& Life Sciences Research Ethics Subcommittee at Maynooth University.

\textbf{Funding} This research was conducted with the financial support of Taighde Éireann – Research Ireland under Grant numbers 18/CRT/6049.

\bibliographystyle{IEEEtran}
\bibliography{references}

\end{document}